\documentclass[preprint]{elsarticle}

\usepackage[utf8]{inputenc}
\usepackage{url}
\usepackage{graphicx}
\usepackage{braket}
\usepackage{float}
\usepackage{array}
\usepackage{gensymb}
\usepackage{amsmath}
\usepackage{physics}
\usepackage{amssymb}
\usepackage{textcomp}
\usepackage{dsfont}
\usepackage{epsfig,amsmath,amsfonts}
\usepackage[hidelinks]{hyperref}
\begin{document}

\begin{frontmatter}

\title{Translating current ALP photon coupling strength bounds to the Randall-Sundrum model}

\author[a]{Shihabul Haque}
\author[a]{Sourov Roy}
\author[a]{Soumitra SenGupta}
 
% \affiliation[a]{School of Physical Sciences, Indian Association for the Cultivation of Science, 2A and 2B Raja S.C. Mullick Road, Kolkata 700032, India}
\affiliation[a]{organization={School of Physical Sciences, Indian Association for the Cultivation of Science},%Department and Organization
            addressline={2A and 2B Raja S.C. Mullick Road}, 
            city={Kolkata},
            postcode={700032}, 
            state={West Bengal},
            country={India}}

%\emailAdd{shihabul1312@gmail.com}
% \emailAdd{tpsr@iacs.res.in}
% \emailAdd{tpssg@iacs.res.in}  

\begin{abstract}
In this article, we look at the current bounds on the coupling strength of axion-like particles (ALPs) with two photons in the context of the Randall-Sundrum (RS)
model. We relate the coupling strength to the compactification radius that governs the size of the extra dimension in the RS warped geometry model 
and show how the current bounds on the ALP can be used to derive appropriate constraints on the size of the extra fifth dimension in the RS model. We show that 
the resulting constraints fail to resolve the gauge hierarchy problem for light/ultralight ALPs and require a massive ALP of at least $m_{a} \gtrsim 0.1$ [GeV] to 
be relevant in the context of the hierarchy problem when the gauge field is in the bulk.
\end{abstract}

% \begin{keyword}
% ALP \sep photons\sep RS model
% \end{keyword}

\end{frontmatter}

\section{Introduction}\label{sect1}
Theories of extra spatial dimensions were originally proposed to achieve electromagnetism and gravity unification through a geometric interpretation. Subsequently,
there was a revival of interest in extra-dimensional models in the context of a possible resolution of the unnatural fine-tuning problem associated with the mass of
the Higgs boson against large radiative corrections which, in turn, have their roots in the large hierarchy between the electroweak and Planck scale. To achieve a
resolution to this problem geometrically, several variants of higher dimensional theories were proposed. Among these, theories with warped extra dimensions drew
special attention because of their inherent property to resolve the gauge hierarchy naturalness problem \cite{naturalness1, naturalness2} without invoking any intermediate scale 
other than the Planck scale (see also \cite{Branchina}). It has also been shown that such types of warped geometry models have their natural origin in an underlying string theory 
which admits of Klebanov-Strassler throat geometry \cite{ks} in the backdrop of an appropriate string compactification. The simplest and most popular warped geometry model was 
proposed by Randall and Sundrum (RS) \cite{RS1999} which portrays the presence of two $3$-branes located at the two orbifold fixed points in a $5$-dimensional 
anti-de Sitter bulk compactified on a $Z_2$ orbifold. This model mimics the essential features of a stringy braneworld model compactified on $\text{AdS}_5 \cross S_5$. Due 
to the anti-de Sitter character of the $5$-dimensional bulk, this model also received special attention in the context of AdS/CFT correspondence \cite{adscft}. \\

The low energy four dimensional effective field theory action of string theory \cite{gsw,pol,senrev1, senrev2}, the supergravity action, contains a massless field, a 
rank two antisymmetric tensor, the Kalb-Ramond (KR) field \cite{kr}. The interpretation of the KR field strength \cite{pmssg} as a torsion in the 
background spacetime \cite{hehl1, hehl2, sabbata} inevitably implies the study of electromagnetism in a spacetime with torsion. It may be observed that the KR field, being a closed 
string excitation, naturally resides in the bulk spacetime. The corresponding three-form field strength of the two-form KR field can be identified with the 
background spacetime torsion. It has also been shown that such models are plagued with the appearance of gauge anomaly and a gauge $U(1)$ Chern-Simons term can 
cancel such anomalies in the background supergravity theory \cite{greenschwarz}. However, the Chern-Simons extension inevitably leads to a gauge-invariant coupling 
of the KR field (or, torsion) with the electromagnetic field \cite{pmssg}. As a result, the dual of the KR field strength in four dimensions, which appears as the
derivative of a scalar field identified with the axion-like particle (ALP), brings in an ALP/photon coupling. Thus, such a coupling is natural in a string inspired 
RS model from the requirement of consistency of the underlying supergravity theory.\\

The 
most important aspect of the RS model is its testability where observable signatures can be searched for both in TeV scale collider physics as well as in various 
astrophysical and cosmological scenarios \cite{rspheno1, rspheno2, rspheno3, rspheno4, rspheno5, rspheno6, rspheno7, rspheno8, rspheno9, rspheno10}. Apart from various scattering and decay processes associated with graviton Kaluza-Klein modes (mass 
$\sim$ TeV) coupled to the standard model fields with inverse TeV couplings \cite{gravkk1, gravkk2}, and the phenomenology of the TeV scale radion (the only extra-dimensional 
modulus in the model) \cite{radion1, radion2, radion3, radion4, radion5}, the RS model also gives rise to an axion phenomenology. The origin of such axions and their coupling with standard model photons 
is due to the coupling of the second-rank massless antisymmetric Kalb-Ramond field with the electromagnetic field (photon). Such a coupling arises from the 
requirement of the Chern-Simons extension to cancel the $U(1)$ gauge anomaly which also restores the $U(1)$ invariance for an appropriate choice of gauge transformation 
of the KR field \cite{axionphoton1, axionphoton2}. The magnitude of the resulting axion-photon coupling is dictated by the value of the warp factor in this model 
\cite{PRD2008, axionphotoncoupling1,axionphotoncoupling2} which, in turn, is determined from the fundamental requirement of the resolution of the gauge hierarchy problem discussed earlier. Whether 
this requirement is consistent with the available constraints on the mass of the axions and its coupling from various experimental findings is the essential 
goal of the present work. In this context, it may be mentioned that there are various proposals of the origin of such couplings in different variants of RS-like models. These can be 
found in references \cite{Bonnefoy2020, Cox2019, Giorgi2024, Flacke2006}. However, the model discussed in this work is a generic one which has its origin in string-inspired low energy supergravity models in the backdrop of a warped 
geometry. We assume, here, that the axion acquires a mass through various non-perturbative mechanisms. In the context of string theory, one such 
mechanism is mass generation through world sheet instanton corrections \cite{wsinstanton1,wsinstanton2}. In our analysis, we consider two distinct scenarios: (a) the gauge 
field propagates in the bulk \cite{gaugebulk1,DAV1999, gaugebulk2, gaugebulk3} and (b) the gauge field is confined on the brane. In both these cases, the KR field, which appears as closed string 
excitation, is assumed to reside in the bulk. In both scenarios, our analysis reveals that resolving the gauge hierarchy problem and satisfying 
the strong experimental bounds of axion-photon coupling simultaneously is strongly disfavoured for ultralight and light axions and allows a window for axion mass $\gtrsim 0.1$ [GeV] for the 
bulk gauge field scenario only. On the other hand, when the gauge field is confined on the brane, the current experimental bounds do not permit a possible 
resolution for the hierarchy problem even up to a few TeV of axion mass. We present our work as follows: \\

We briefly describe the RS model and the origin of the axion-photon coupling in the RS model. Subsequently, we highlight the various bounds set by different 
experiments on the axion-photon coupling and the corresponding axion mass. This is then compared to test the viability of the RS model in the context of the 
resolution of the fine-tuning/gauge hierarchy problem as discussed earlier.

\section{Axion Photon coupling in the Randall-Sundrum model}\label{sect2}
The metric ansatz for the RS model is given by \cite{RS1999}, 
\begin{equation}
    ds^{2} = e^{-2 \sigma(y)} \eta_{\alpha \beta} dx^{\alpha} dx^{\beta} + r^{2}_{C} d \phi^{2}
\end{equation}
The RS model describes a spacetime with five dimensions with the fifth dimension warping the other four. This warp factor is captured by the initial exponential
term where $\sigma (y) = k r_{C} \phi = k y$. Here, $\phi$ denotes the coordinate of the extra dimension and, $r_{C}$, its radius while $k$ is related to the five 
dimensional Planck scale as,
\begin{equation}
    k = \sqrt{-\frac{\Lambda}{24M^{3}}}
\end{equation}
Where $\Lambda$ is the bulk cosmological constant which is negative for an anti-de Sitter bulk and, $M$ is the five dimensional Planck mass. In the RS model, the five dimensional
cosmological constant is assumed to be of the order of $M^5$ which results in the value of $k$ to be of the order of $M$. Using the relation between the four and five dimensional
Planck scales in the RS model, one finds that $k$ is of the order of $M_p$.
The model describes a bulk spacetime with two $3$-branes at $\phi = 0, \ \pi$. The visible universe is located on the latter location while 
the former is hidden. There is a $\mathbb{Z}_{2}$ symmetry in the sense that the regions of $\phi \in [-\pi, 0]$ and $\phi \in [0, \pi]$ are identified. 
\begin{figure}[H]
    \centering
        \includegraphics[width=0.8\textwidth]{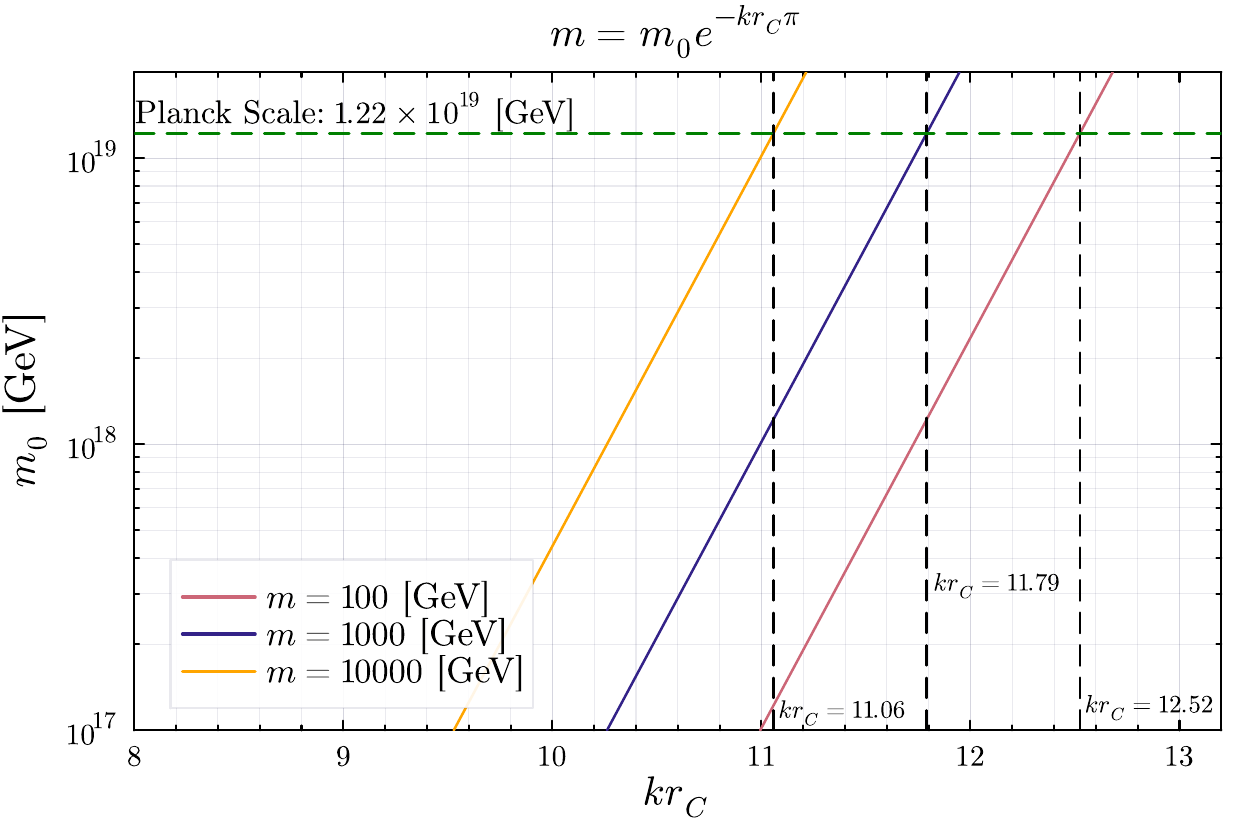} 
        \caption{Suppression from the Planck scale in the RS model. The coloured solid lines depict the constant $m$ curves in this plane given by eqn. \eqref{mass}. The dotted green
        line indicates the Planck scale. The intersection of the constant $m$ curves with the Planck scale gives us the value of $k r_C$ required for a suppression
        from the Planck scale to $m$.}
        \label{fig1}
\end{figure}
The Randall-Sundrum model utilises the presence of an extra dimension with an exponential warp factor to present a potential solution to the hierarchy problem in
particle physics. If the energy scale on the hidden brane is $m_{0}$, then that scale is suppressed to $m$ on the visible brane where, 
\begin{equation}\label{mass}
    m = m_{0}e^{-kr_{C}\pi}
\end{equation}
The compactification factor in the model, $k r_{C}$, has to take a value of around $11.79$ to cause a suppression from the Planck scale to the TeV scale
\cite{PRD2008}. We illustrate this idea in Fig. \ref{fig1} by plotting $m$ (as coloured lines) in the ($m_0$, $k r_C$) plane. The intersection of these curves with the Planck 
scale (the dotted green line) gives us the value of $k r_C$ leading to the desired suppression. We see that the value of $kr_C$ corresponding to $11.79$ leads to a suppression from the Planck 
scale to around the TeV scale while a value of $11.06$ gives a weaker, partial suppression up to a $10$ TeV scale. Higher suppression, for instance to the electroweak scale, would 
require a higher value of the compactification parameter. In our analysis, it will be observed that the value of $kr_C$ determines various bounds. As discussed earlier, the value of $k$ is assumed to be of the order of $M_p$ and $r_C$
of the order the Planck length, $L_p$, such that $kr_C$ is $\mathcal{O}(10)$. This implies that $c = M_p r_C$ is also an $\mathcal{O}(10)$ number. The choice of the values of
$k$ and $r_C$ close to Planck mass and Planck length are the hallmark of RS model as it guarantees that without introducing any significant hierarchy in the parameters of the model,
we can produce a large hierarchy between the Planck scale and the TeV scale. In addition, the choice of $k$ and $r_C$ close to the Planck scale (quantum gravity scale) guarantees
that these parameters are stable against radiative corrections. Therefore, if we want to have the Planck to TeV scale warping and also want to keep the parameters of the model
close to Planck scale without creating any hierarchy between them, then $kr_C \sim M_p r_C \sim \mathcal{O}(10)$ is an optimal choice. By relating the 
axion/photon coupling strength to the compactification factor, we can translate our current experimental 
observations on the coupling strength to bounds on the size of the extra dimension giving us the opportunity to review the kind of situations under which the 
Randall-Sundrum model can address the hierarchy issue based on our present observational data. \\

The effective four dimensional action we consider after compactification of the extra dimension is given by, 
\begin{equation}\label{action_original}
    \mathcal{S} = \int d^{4} x \sqrt{-g} \Big[M^{2}_{p}\mathcal{R} - \frac{1}{12} \Bar{H}^{\mu\nu\rho}\Bar{H}_{\mu\nu\rho} - \frac{1}{4}F^{\mu\nu}F_{\mu\nu} \Big]
\end{equation}
Here, $F^{\mu\nu}$ is the usual Maxwell field, $\mathcal{R}$ is the Ricci scalar and, 
\begin{equation}\label{anomaly}
    \Bar{H}_{\mu\nu\rho} = \partial_{[\mu} B_{\nu\rho]} + \frac{\beta}{M_{p}} A_{[\mu} F_{\nu\rho]}
\end{equation}
$B_{\mu\nu}$ is the Kalb-Ramond (KR) field.
The origin of the second term in eqn. \eqref{anomaly} lies in the well-known Green-Schwarz anomaly cancellation mechanism, both in the context of gauge as well as gravitational anomaly 
\cite{gsw, greenschwarz, gsw2, Alvarez-Gaume}. The coefficient $\beta$ contains contributions from both the compactification procedure and the anomaly cancellation requirement. We now
briefly describe the origin of the KR-photon interaction in this model below. In the context of $\text{U}(1)$ gauge anomaly, a Chern-Simons $3$-form is added to the field
strength $H_{\mu\nu\rho}$ of the $2$-form gauge fixed $B_{\mu\nu}$ (where $H_{\mu\nu\rho} = \partial_{[\mu} B_{\nu\rho]}$) as,
\begin{equation}
H_{\mu\nu\rho}\ \rightarrow \ \Bar{H}_{\mu\nu\rho} = H_{\mu\nu\rho} + \frac{\beta}{M_{p}} A_{[\mu} F_{\nu\rho]}
\end{equation}
This makes the theory free from $\text{U}(1)$ gauge anomaly. The modified tensor, $\tilde{H}_{\mu\nu\rho}$, is gauge invariant when the KR field transforms under a 
$\text{U}(1)$ gauge transformation as $\delta B_{\mu\nu} = - \omega F_{\mu\nu}$. \\

The four dimensional gauge invariant action is given as, 
\begin{equation}
\mathcal{S} = \int d^4 x \sqrt{-g} \left[M^{2}_{p}\mathcal{R} - \frac{1}{4} F^{\mu\nu} F_{\mu\nu} - \frac{1}{12}
\Bar{H}^{\mu\nu\rho}\Bar{H}_{\mu\nu\rho}\right]
\end{equation}
Using the expression of $\Bar{H}_{\mu\nu\rho}$ as described above, we obtain the leading order interaction term, 
\begin{equation}
\mathcal{S}_{int} = \frac{\beta}{M_{p}} \int d^4 x \sqrt{-g}\ H^{\mu\nu\rho}A_{[\mu} F_{\nu\rho]}
\end{equation}

It may also be noted that the KR field strength is invariant under the KR gauge transformation, 
$\delta B_{\mu\nu} = \partial_{[\mu} A_{\nu]}$, and also satisfies the Bianchi identity.  
The above analysis can also be extended to higher dimensions which after appropriate compactification yields the effective field equations in four dimensions, as,
\begin{equation}
    D_{\mu} \Bar{H}^{\mu\nu\rho} = 0, \ D_{\mu} F^{\mu\nu} = \frac{\beta}{M_{p}} \Bar{H}^{\nu\rho\lambda}F_{\rho\lambda} 
\end{equation}
An exact expression for the compactification parameter, $\beta$, will be derived in detail in the following subsections. Now, in four dimensions, we can express 
$H^{\mu\nu\rho}$ in terms of a massless scalar field, $a$, as, 
\begin{equation}
    H^{\mu\nu\rho} = \epsilon^{\mu\nu\rho\lambda} \partial_{\lambda} a 
\end{equation}
This is the axion field. In terms of the axion, the field equations become, 
\begin{equation}
    \partial^{\mu} \partial_{\mu} a = \frac{\beta}{2 M_{p}} F^{\mu\nu} \Tilde{F}_{\mu\nu}, \ D_{\mu} F^{\mu\nu} = - \frac{2\beta}{M_{p}} 
    (\partial_{\mu} a) \Tilde{F}^{\mu\nu}
\end{equation}
The effective four dimensional action then becomes, 
\begin{equation}\label{interactionalp}
    \mathcal{S} = \int d^{4} x \sqrt{-g} \Big[M^{2}_{p}\mathcal{R} - \frac{1}{2} (\partial_{\mu} a \ \partial^{\mu} a - m_{a}^{2}a^{2}) - 
    \frac{\beta}{2 M_{p}} a F^{\mu\nu} \Tilde{F}_{\mu\nu} - \frac{1}{4}F^{\mu\nu}F_{\mu\nu} \Big]
\end{equation}
For our present discussion, an ALP can gain a mass term
due to various contributions - here, we have inserted a mass term by hand without focusing on its origins. We briefly outline one such mass generation mechanism for the axion in our proposed $5$D warped spacetime (for further details, see \cite{Gherghetta2020, Csaki2019}). Considering a QCD gauge boson in the bulk alongside a 
$U(1)$ gauge boson and identifying the zero mode of the fifth component of the $U(1)$ gauge boson with the axion, it is shown that the anomaly cancelling bulk Chern-Simons term 
upon compactification of the $5^{\text{th}}$ dimension leads to an axion mass term through the instanton corrections in the model. In the limit when the theory is strongly coupled 
at $\Lambda_5$ ($5$D cutoff scale of the theory below the strong coupling scale), the axion mass is given by, 
\begin{equation}
    m^2_a \approx \frac{\Lambda^4_5}{f^2} \text{ with } \frac{1}{f} = b_{\rm cs} g_s L
\end{equation}
Here, $b_{\rm cs}$ is a dimensionless constant relating the coupling of the $U(1)$ gauge boson to the QCD gauge boson, $g_s$ is the four dimensional QCD gauge coupling related to
$g_5$ as $g^2_s = g^2_5/L$ and $L = \pi R$, where $R$ is the radius of the extra dimension. A large contribution to the axion mass can be obtained close to the non-perturbative limit 
through small instanton corrections. The presence of the extra dimensions adds to a UV modification of QCD at the scale $1/R >> \Lambda_{\rm QCD}$ ($R$ is the compact radius and $\Lambda_{\rm QCD} \approx 300$ MeV is the 
QCD scale). Instantons of size $\rho < R$ (small instantons) generate large contributions to the axion mass. \\

Several consequences of the axion-photon interaction in eqn. 
\eqref{interactionalp} in the context of braneworld physics or otherwise have been discussed in literature \cite{axionphoton1, axionphoton2,axionphotoncoupling1,PRD2008}. Generically,
we expect a further interaction term of the form \cite{Capanelli2023},

\begin{equation}
    \mathcal{L}_{\rm int} = \sum_i g_{a f_i f_i} \frac{{\partial_\mu} a}{f_a} \bar{f}_i \gamma_\mu \gamma_5 f_i
\end{equation}
This contributes to the effective ALP/photon coupling at one loop level via an additional term of the form \cite{Bauer2017},
\begin{equation}
g^{\rm eff}_{a \gamma \gamma} = g_{a \gamma \gamma} + \sum_i \frac{N_c^{f_i} Q_{f_i}^2}{8 \pi^2} g_{a f_i f_i} B_1 (\tau_{f_i}),
\end{equation}
The fermion loop functions can be evaluated to be $B_1(\tau_f) \approx 1$ for $m_f \ll m_a$, while 
$B_1 (\tau_f) \approx - m^2_a/12 m^2_f$ for $m_f \gg m_a$. Thus we see that the fermions heavier than the ALP decouple whereas the fermions lighter than the ALP adds a 
contribution of order $g_{a f_i f_i}/8 \pi^2$ to the effective Wilson coefficient $g^{\rm eff}_{a \gamma \gamma},$ which is suppressed. \\

On the other hand, the derivative couplings can be removed by performing an ALP-dependent chiral transformation of the fermion fields,
\begin{equation}
    f_i \rightarrow e^{i\frac{g_{a f_i f_i}}{f_a} a \gamma_5} f_i, \quad \bar{f}_i \rightarrow \bar{f}_i e^{i\frac{g_{a f_i f_i}}{f_a} a \gamma_5}
\end{equation}
The fermion mass terms become complex under this redefinition, involving $e^{2i (g_{a f_i f_i}/f_a) a \gamma_5}$. However, since this transformation is anomalous, it also gives rise to 
a contribution to the ALP/photon coupling \cite{Eberhart:2025lyu} in the interaction Lagrangian which, up to $\mathcal{O}(a/f_a)$, is given by,
\begin{equation}
    \mathcal{L}_{\rm int} = -i a \sum_i \frac{2 m_{f_i}}{f_a} g_{a f_i f_i} \bar{f}_i \gamma_5 f_i + \frac{\alpha}{4\pi} \frac{a}{f_a} \sum_i 2 N_c^{f_i} {\tilde Q}_{f_i}^2 
    g_{a f_i f_i} F^{\mu\nu} {\tilde F}_{\mu \nu}
\end{equation}
where ${\tilde Q}_{f_i} = Q_{f_i} / e$ and $\alpha = \frac{e^2}{4\pi}$. This leads us to the same $g_{a\gamma \gamma}^{\rm eff}$ given above. Therefore, the total
effective coupling due to the fermions is unchanged and suppressed. Thus, we neglect this contribution in our
current analysis. \\

As we can see, the parameter $\beta$ now enters into the above action as the ALP/photon coupling strength, implying that the same factor controls the extra-dimensional information 
and regulates ALP/photon mixing. This allows us to relate one to the other - but before that, we need to determine the exact relation of $\beta$ to the extra-dimensional 
parameters. We shall now look at two separate cases - when the gauge field is present on the bulk and when it is restricted to the brane only. 

\subsection{Gauge field in the bulk}
In order to derive the expression for the ALP/photon coupling strength in terms of the parameters in the RS model, we must begin from the KR action in five 
dimensions \cite{PRD2008, DAV1999}, 
\begin{equation}
    \mathcal{S}_{KR} = -\frac{1}{12} \int d^{5}x \ \sqrt{-G} \ H_{MNL}H^{MNL}
\end{equation}
The metric determinant is given as $\sqrt{-G} = e^{-4kr_{C}\phi}r_{C}$. Using the invariance of the KR action under the gauge transformation, $\delta
B_{MN} = \partial_{[M}A_{N]}$, we set $B_{4\mu}$ to $0$ (here $\mu$ runs from $0 - 3$). This gives us, 
\begin{equation}
    \mathcal{S}_{KR} = -\frac{1}{12} \int d^{4}x \int d\phi \ r_{C} e^{2kr_{C}\phi}\Big[\eta^{\mu\alpha}\eta^{\nu\beta}\eta^{\gamma\lambda}H_{\mu\nu\lambda}
    H_{\alpha\beta\gamma} - \frac{3e^{-2kr_{C}\phi}}{r^{2}_{C}}\eta^{\mu\alpha}\eta^{\nu\beta}B_{\mu\nu}\partial^{2}_{\phi}B_{\alpha\beta} \Big] 
\end{equation} 
Similarly, we start from the five dimensional Maxwellian part, 
\begin{equation}
    \mathcal{S}_{EM} = -\frac{1}{4} \int d^{5}x \ \sqrt{-G} \ F_{MN}F^{MN}
\end{equation}
We use the gauge freedom to set $A_{4} = 0$. Proceeding as before, we get, 
\begin{equation}
    \mathcal{S}_{EM} = -\frac{1}{4} \int d^{4}x \int d\phi \ r_{C} \Big[\eta^{\mu\alpha}\eta^{\nu\beta}F_{\mu\nu}F_{\alpha\beta} - \frac{2}{r^{2}_{C}}
    \eta^{\mu\alpha}A_{\mu}\partial_{\phi}(e^{-2kr_{C}\phi}\partial_{\phi}A_{\alpha}) \Big]
\end{equation}
We use standard Kaluza - Klein decomposition for the KR and the electromagnetic fields, 
\begin{equation}
    B_{\mu\nu}(x^{\mu}, \phi) = \frac{1}{\sqrt{r_{C}}}\sum_{n=0}^{\infty} B^{n}_{\mu\nu}(x^{\mu})\chi^{n}(\phi),\
    A_{\mu}(x^{\mu}, \phi) = \frac{1}{\sqrt{r_{C}}}\sum_{n=0}^{\infty} A^{n}_{\mu}(x^{\mu})\xi^{n}(\phi)
\end{equation}
The zero mode solutions thus obtained are, 
\begin{equation}
    \chi^{0} = \sqrt{kr_{C}}e^{-kr_{C}\pi},\ \xi^{0} = 1/\sqrt{2\pi}
\end{equation}
We now simplify the coupling term in a similar manner. The KR-photon interaction term before compactification is given by,
\begin{equation}
    \mathcal{S}_{int} = -\frac{1}{12 M^{3/2}} \int d^{5}x \ \sqrt{-G} \ H^{MNL} A_{[M}F_{NL]}
\end{equation}
We now determine the expression of $\beta$ by implementing RS compactification. This gives us, 
\begin{equation}
    \mathcal{S}_{int} = -\frac{1}{12 M_{p}^{3/2}} \int d^{4}x \int d\phi \ \Big[r_{C} e^{2kr_{C}\phi}\eta^{\mu\alpha}\eta^{\nu\beta}\eta^{\gamma\lambda}H_{\mu\nu\lambda}
    A_{[\alpha}F_{\beta\gamma]} + \frac{6}{r_{C}}\eta^{\mu\alpha}\eta^{\nu\beta}A_{\beta}(\partial_{\phi}A_{\alpha})(\partial_{\phi}B_{\mu\nu}) \Big]
\end{equation}
Using the Kaluza - Klein decomposition and substituting the zero mode solutions, we find, 
\begin{equation}
    \mathcal{S}_{int} = -\frac{1}{12}\int d^{4}x \Bigg[\int d\phi \ \frac{e^{2kr_{C}\phi}}{\sqrt{r_{C}M^{3}_{p}}} \chi^{0} (\xi^{0})^{2}\Bigg] H^{\mu\nu\lambda}
    A_{[\mu}F_{\nu\lambda]} \equiv -\frac{1}{12}\int d^{4}x \frac{\beta}{M_{p}} H^{\mu\nu\lambda}
    A_{[\mu}F_{\nu\lambda]}
\end{equation}
This implies that the parameter $\beta$ is related to the radius of the fifth dimension as, 
\begin{equation}
    \beta = \int d\phi \ \frac{e^{2kr_{C}\phi}}{\sqrt{r_{C}M_{p}}} \chi^{0} (\xi^{0})^{2}= \frac{1}{2\pi k r_{C}}\sqrt{\frac{k}{M_{p}}} e^{k r_{C} \pi}
\end{equation}
% Therefore, the parameter $\beta$ is related to the radius of the fifth dimension as, 
% \begin{equation}
%     \beta = \frac{1}{2\pi k r_{C}}\sqrt{\frac{k}{M_{p}}} e^{k r_{C} \pi}
% \end{equation}
Therefore, the usual axion/photon coupling (denoted by $g_{a \gamma\gamma}/4$ or $1/4M$) can be expressed as, 
\begin{equation}
    \frac{1}{4} g_{a \gamma\gamma} = \frac{\beta}{2 M_{p}} %\Rightarrow \beta = \frac{g_{a\gamma\gamma} M_{p}}{2}
\end{equation} 
Assuming that $k \sim M_{p}$, we have, 
\begin{equation} \label{krc_bulk}
    g_{a\gamma\gamma} = \frac{2}{M_{p}} \beta \approx \frac{1}{\pi k r_{C} M_{p}}e^{kr_{C}\pi}
\end{equation}
We now do the same for the case of gauge fields on the brane.
\subsection{Gauge field on the brane}
When the electromagnetic fields are assumed to be confined on the flat visible 3-brane (at $\phi=\pi$), the Maxwell part of the action in eqn. 
\eqref{action_original} becomes \cite{DM_2004}, 
\begin{equation}
    \mathcal{S}_{EM} = -\frac{1}{4} \int d^{5}x \ \sqrt{-G} \ F_{\mu\nu}F^{\mu\nu} \delta(\phi-\pi) = -\frac{1}{4} \int d^{4}x \ \sqrt{-g_{vis}} \ 
    g^{\mu\alpha}g^{\nu\beta} F_{\mu\nu}F_{\alpha\beta} 
\end{equation}
We have $\sqrt{-g_{vis}} = e^{-4kr_{C}\pi}$, thus, 
\begin{equation}
    \mathcal{S}_{EM} = -\frac{1}{4} \int d^{4}x \ \eta^{\mu\alpha}\eta^{\nu\beta} F_{\mu\nu}F_{\alpha\beta}
\end{equation}
The calculations for the KR field remain the same as before. The interaction portion of the action now becomes,
\begin{equation}\nonumber
    \mathcal{S}_{int} = -\frac{1}{12 M^{1/2}} \int d^{5}x \ \sqrt{-G} \ \delta^{\mu}_{M}\delta^{\nu}_{N}\delta^{\lambda}_{L} H^{MNL} A_{[\mu}F_{\nu\lambda]}
    \delta(\phi-\pi) 
\end{equation}
% \begin{equation}
%     = -\frac{1}{12 M^{1/2}} \int d^{4}x \int d\phi \ e^{2kr_{C}\phi} r_{C} \ \eta^{\mu\alpha}\eta^{\nu\beta}\eta^{\gamma\lambda} H_{\alpha\beta\gamma} A_{[\mu}
%     F_{\nu\lambda]}\delta(\phi-\pi)
% \end{equation}
Substituting the zero mode solution for the KR field and integrating out the extra dimensional coordinate, 
% \begin{equation}
%     \mathcal{S}_{int} = -\frac{1}{12 M^{1/2}} \int d^{4}x \int d\phi \ e^{2kr_{C}\phi} r_{C} \ \eta^{\mu\alpha}\eta^{\nu\beta}\eta^{\gamma\lambda} 
%     \frac{1}{\sqrt{r_{C}}} \chi^{0}H_{\alpha\beta\gamma}^{0} A_{[\mu}F_{\nu\lambda]}\delta(\phi-\pi)
% \end{equation}
% Or, 
\begin{equation}
    \mathcal{S}_{int} = -\frac{1}{12} \int d^{4}x \ \Bigg[\sqrt{\frac{k}{M_{p}}}r_{C} e^{kr_{C}\pi}\Bigg] \eta^{\mu\alpha}\eta^{\nu\beta}\eta^{\gamma\lambda}
    H_{\alpha\beta\gamma}^{0} A_{[\mu}
    F_{\nu\lambda]}
\end{equation}
This implies, 
\begin{equation}
    \beta = r_{C} M_{p}\sqrt{\frac{k}{M_{p}}} e^{kr_{C}\pi}\ \approx\ c\ \sqrt{\frac{k}{M_{p}}} e^{kr_{C}\pi}
\end{equation}
Here, $c\ (=r_{C}M_{p})$ is an $\mathcal{O}(10)$ number (since we assume that $k\sim M_{p}$ with $kr_{C}$ around $11.79$). The axion/photon coupling 
can be expressed as, 
\begin{equation}
    \frac{1}{4} g_{a \gamma\gamma} = \frac{\beta}{2 M_{p}} 
\end{equation} 
Again, assuming that $k \sim M_{p}$, we have, 
\begin{equation} \label{krc_brane}
    g_{a\gamma\gamma} = \frac{2}{M_{p}} \beta \approx \frac{2c}{M_{p}}e^{kr_{C}\pi}
\end{equation}
Currently, we have data on the bounds of $g_{a\gamma\gamma}$ from various experiments and observations, both collider based and from astrophysics and cosmology. 
Using eqns. \eqref{krc_bulk} and \eqref{krc_brane}, we can translate those bounds on the coupling strength to bounds on the compactification factor in the 
RS model. This is done in the following section. 

\section{Analysis of current bounds}\label{sect3}
\begin{figure}[H]
    \centering
        \includegraphics[width=\textwidth]{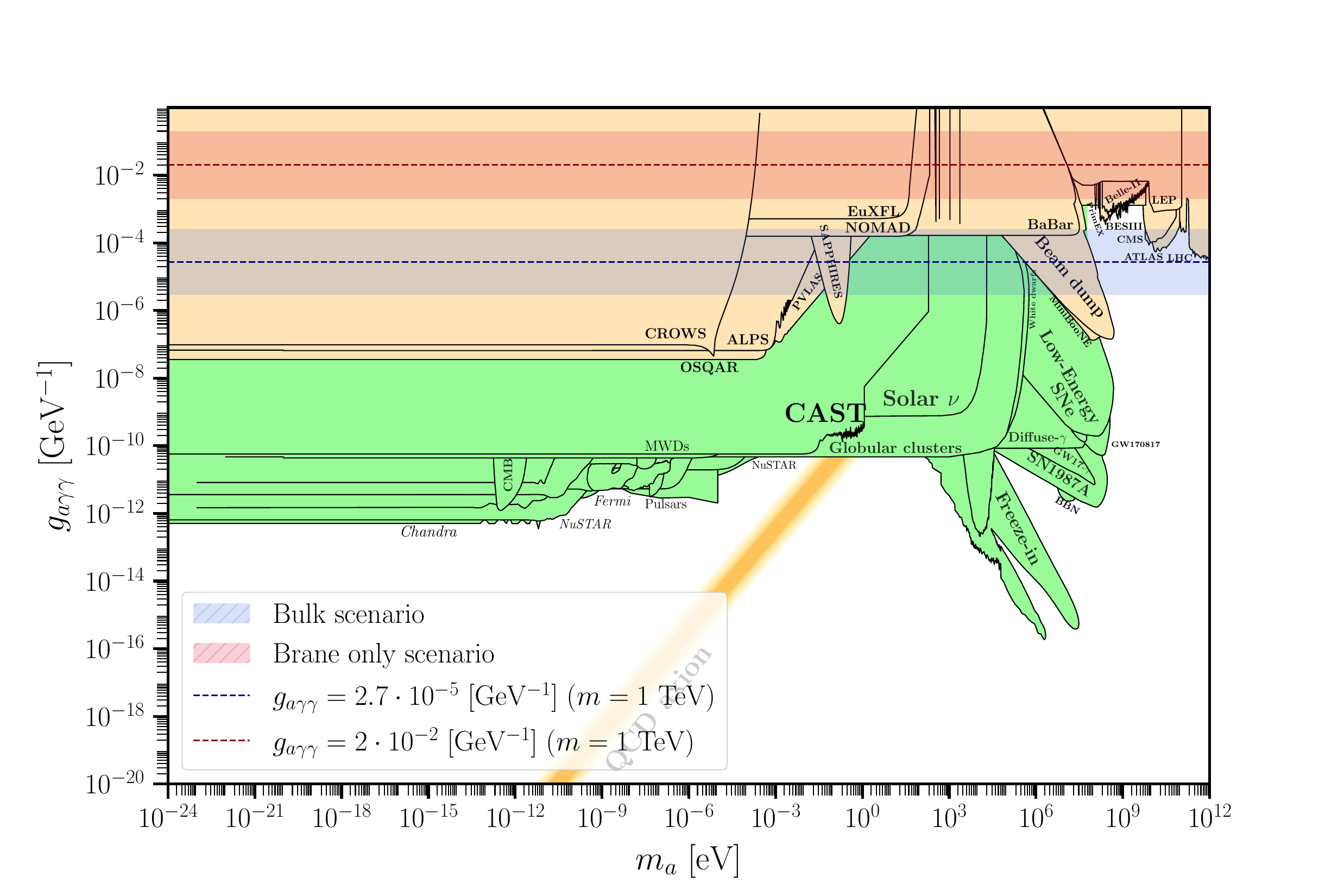} 
        \caption{Existing bounds on ALP/photon coupling strength. Bounds from astrophysics are given in green while those from colliders and 
        beam dump experiments are given in yellow. The QCD axion line is shown in gold. The shaded bands represent the estimated range of coupling strengths for suppression to 
        $0.1$ [TeV] (top edge) to $10$ [TeV] (bottom edge) in the RS model. The dotted lines within the bands represent coupling strengths for suppression to the TeV scale.
        Note that the coupling to the photons is set to $g_{a\gamma\gamma}/4$ which, in our model, is equal to $\beta/2M_p$ and the data is drawn from
        \cite{CJHAxionLimits} for this and the following figures.}
        \label{fig2}
\end{figure}

Let us determine the ALP/photon coupling we need for $k r_{C} = 11.79$ (that is, suppression to the TeV scale). The Planck mass is taken to be $1.22 \times 10^{19}$ [GeV]. For the bulk case, this 
gives us (using eqn. \eqref{mass} with $m_0 = M_p$), 
\begin{equation}\label{critg_bulk}
    g_{a\gamma\gamma} = \frac{1}{\pi k r_{C} M_{p}}e^{kr_{C}\pi} = \frac{1}{m \ln(M_p/m)}
\end{equation}
For suppression to the TeV scale ($m = 1$ [TeV]), we find,
\begin{equation}
    g_{a\gamma\gamma} \approx 2.7 \times 10^{-5} \ [\text{GeV}^{-1}]
\end{equation}
On the other hand, for the case of the gauge field confined to the visible $3$-brane, we have, 
\begin{equation}\label{critg_brane}
    g_{a\gamma\gamma} = \frac{2c}{M_{p}}e^{kr_{C}\pi} = \frac{2c}{M_p} \left(\frac{M_p}{m}\right) \xrightarrow{m \ =\ 1\ \text{[TeV]}} 2c \times 10^{-3} \ [\text{GeV}^{-1}]
\end{equation}
This results, alongside current observational bounds on the ALP/photon coupling strength, are depicted in Fig. \ref{fig2} based on data from various experiments collected and updated at
\cite{CJHAxionLimits}. The lightly shaded bands correspond to the coupling strengths estimated from the RS model for suppression from the Planck scale to the range $0.1$ to $10$ [TeV]. The
upper edge of the bands correspond to $0.1$ [TeV] while the lower edge corresponds to $10$ [TeV]. The dotted lines within the bands represent coupling strengths for suppression to the 
TeV scale (as calculated above). 
The bounds are broadly astrophysical in nature or those coming from laboratory or collider experiments (excluding axion dark matter searches). This 
includes those imposed by the CERN Axion Solar Telescope (CAST) for ALPs produced in the Sun's core \cite{CAST},  
from globular clusters \cite{globularcluster1,globularcluster2} and irreducible freeze-in considerations \cite{Irrbackground}. Supernovae explosions also
provide excellent scope for studying the radiative decay of ALPs through the $a\rightarrow\gamma\gamma$ process. The plot includes limits 
derived from the decay of ALPs from the supernova SN1987A into photons \cite{SN1987_1,SN1987_2} (see also \cite{Damiano2}). The KSVZ and DFSZ QCD axion models 
\cite{KSVZ1,KSVZ2,DFSZ1,DFSZ2} are shown in gold. Recently, reference \cite{Candon} also imposed newer bounds based on the radiative decay of ALPs from the M$82$ galaxy. 
The constraints for heavier ALPs are described later. \\

The derived values are already quite large compared to current bounds - particularly so for the brane case. Fig. \ref{fig2} clearly shows us that such 
coupling strengths can only be possible for heavy ALPs with $m_{a} \gtrsim 0.1$ [GeV] in the bulk case while masses upto a few TeV are ruled out for the brane case.
Lighter ALPs seem to be strongly disfavoured. Therefore, for the RS model to be relevant in the context of the hierarchy issue, we can only have heavy ALPs 
in our universe. 
\begin{figure}[H]
    \centering
        \includegraphics[width=\textwidth]{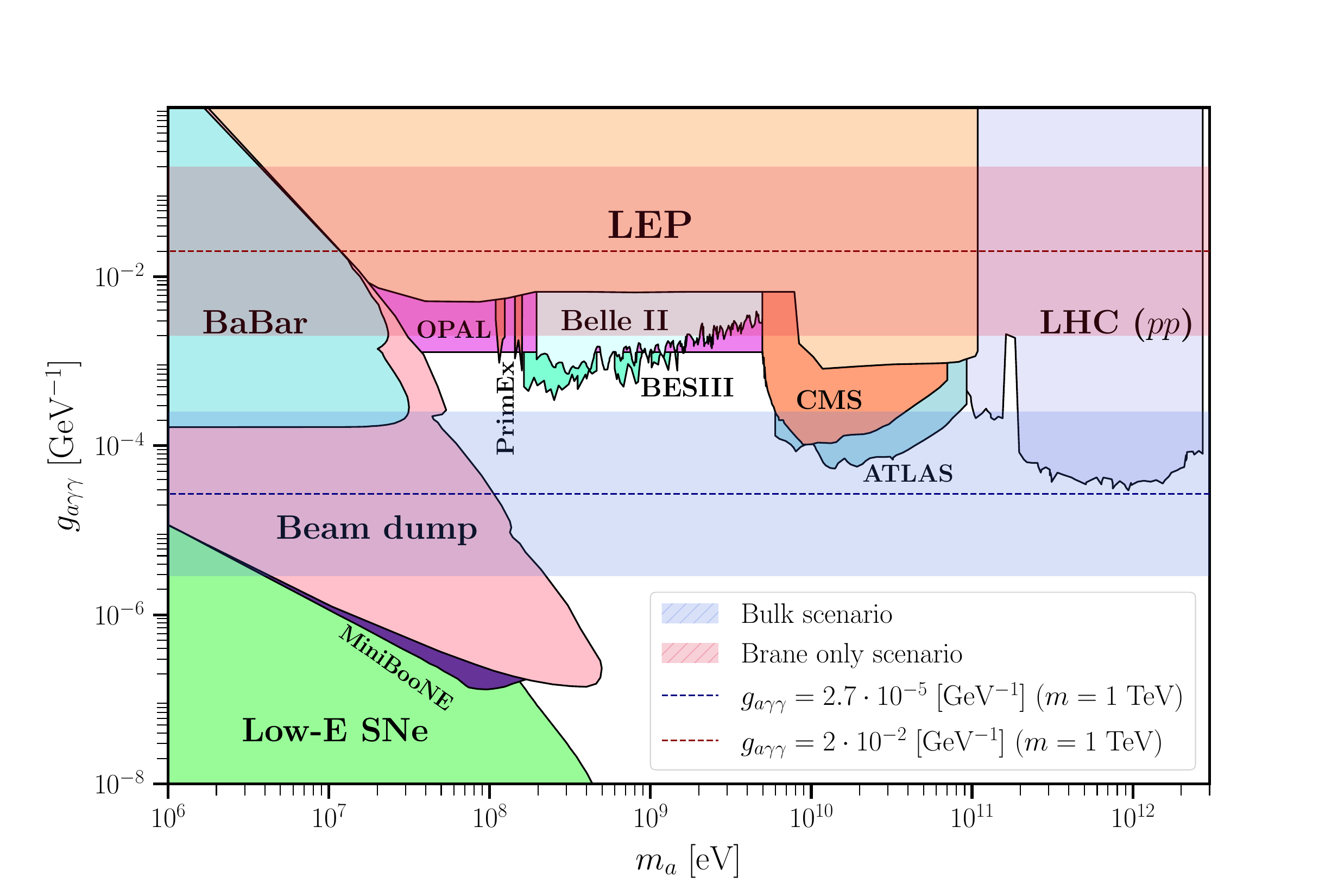} 
        \caption{Bounds on $g_{a\gamma\gamma}$ for heavy ALPs. The shaded bands represent the estimated range of coupling strengths for suppression to 
        $0.1$ [TeV] (top edge) to $10$ [TeV] (bottom edge) in the RS model. The dotted lines within the bands represent coupling strengths for suppression to the TeV scale.}
        \label{fig3}
\end{figure}
Bounds on $g_{a\gamma\gamma}$ for heavy ALPs with $m_{a} > 10^{6}$ [eV] as per eqns. \eqref{critg_bulk} and \eqref{critg_brane} are shown in Fig. \ref{fig3}. This 
includes data from light by light scattering in Pb-Pb collisions performed at the ATLAS experiment \cite{ATLAS}, dark photon searches at the BaBar experiment \cite{BaBar} and searches for the 
$\gamma\gamma\rightarrow a \rightarrow\gamma\gamma$ process in ultra-peripheral Pb-Pb collisions at the CMS experimental collaboration \cite{CMS}. Bounds were also given by 
the decays of $J/\psi \rightarrow \gamma a;\ a\rightarrow \gamma\gamma$ at BESIII collaboration \cite{BESIII} and $e^{+}e^{-}\rightarrow \gamma a;\ 
a\rightarrow \gamma\gamma$ at the Belle II experiment \cite{BelleII}. Beam dump searches, which involve probing radiative decay of ALPs, include runs at the CHARM experimental collaboration \cite{BD1}, 
SLAC \cite{BD2, BD3} and others including the NA64 collaboration at CERN \cite{BD4, BD5, BD6, BD7} and the MiniBooNE collaboration \cite{MiniBooNE}. ALP decays into 
$2\gamma,\ 3\gamma$ were investigated in the OPAL experiment \cite{OPAL,OPAL2}. Searches using the $\pi^{0}\rightarrow\gamma\gamma$ decay process were carried out using the PrimEx 
data \cite{PrimEx1, PrimEx2}. Constraints using low explosion energy supernovae \cite{Sne}, X-ray observations of the neutron star merger GW170817 \cite{GW170817},
and LEP and LHC limits, including searches based on radiative decay of ALPs and ALP couplings to Z bosons \cite{LEP}, are also shown. For more discussion on the 
various bounds, see \cite{Bauer2017,Mimasu2015,Caputo2024}.
\begin{figure}[H]
    \centering
        \includegraphics[width=\textwidth]{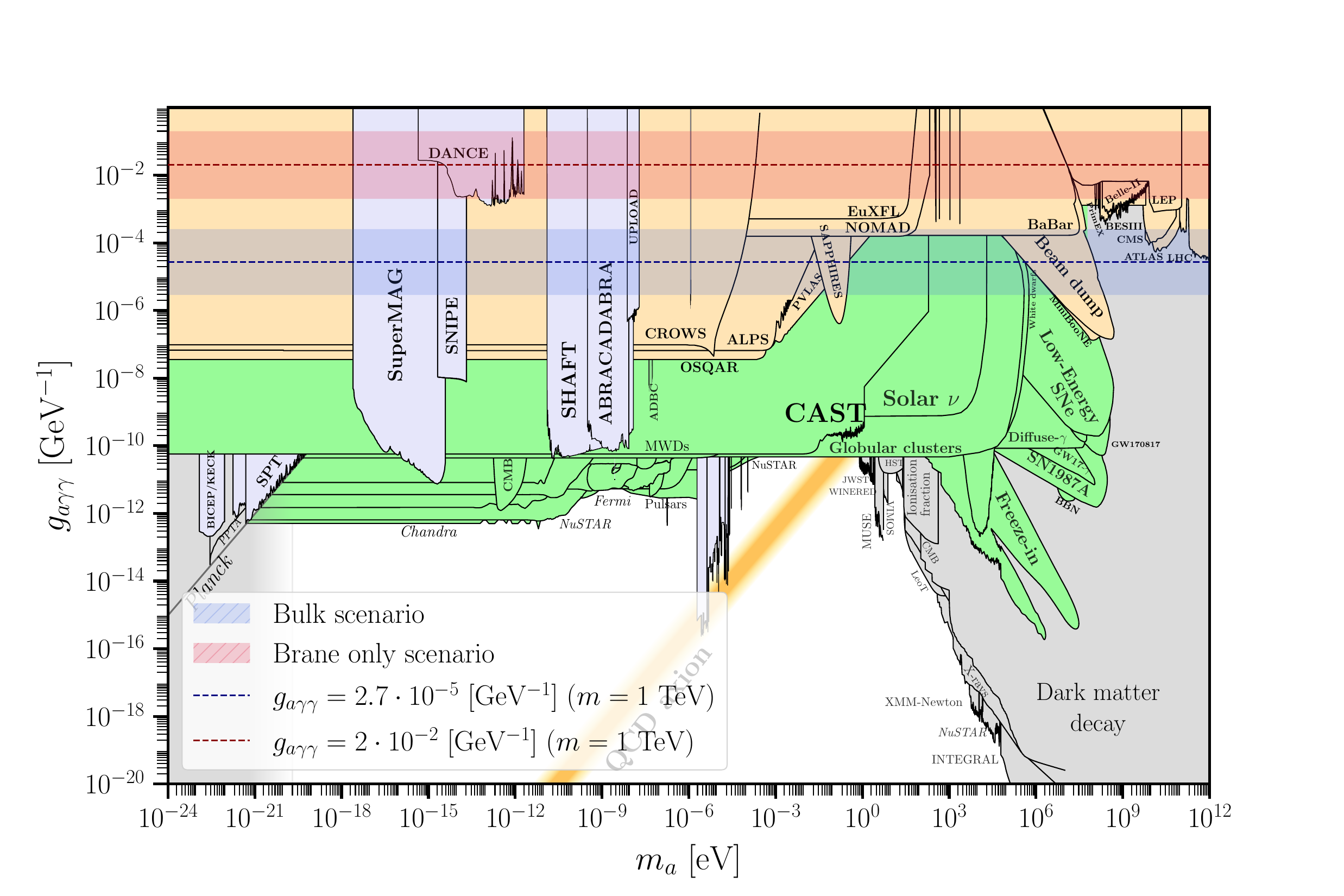} 
        \caption{Existing bounds on ALP/photon coupling strength including DM searches. Bounds from DM searches in colliders or the like are given in lavender while 
        those from astrophysical DM searches are given in gray. The shaded bands represent the estimated range of coupling strengths for suppression to 
        $0.1$ [TeV] (top edge) to $10$ [TeV] (bottom edge) in the RS model. The dotted lines within the bands represent coupling strengths for suppression to the TeV scale.}
        \label{fig4}
\end{figure}
More stringent conclusions can be drawn if ALPs are also considered to be dark matter candidates and appropriate bounds are imposed as shown in Fig. \ref{fig4}. The figure includes bounds from ALP DM searches based on
expected interactions with magnetic fields at the ABRACADABRA \cite{ABRACADABRA1,ABRACADABRA2}, SHAFT \cite{SHAFT}, SuperMAG \cite{Supermag1,Supermag2} and SNIPE
\cite{SNIPE} experiments as well as bounds from birefringence and polarization based searches at the ADBC \cite{ADBC} and DANCE (first run) \cite{DANCE} experiments.
The plot also includes bounds from CMB based dark matter searches \cite{Capozzi_2023} and considerations from DM decay lifetimes. \\

It is clear from Figs. \ref{fig2}-\ref{fig4} that the ALP masses are quite heavily restricted - the allowed ranges include $m_{a} \gtrsim 0.1$ [GeV] 
for the bulk case and even higher for the brane case while the ALP DM derived from such considerations also seems to be strongly disfavoured.

\section{Summary and discussion}\label{sect4}

We have translated the current bounds on the ALP/photon coupling strength to bounds on the compactification radius in the Randall-Sundrum model of spacetime. We have
shown that for the Randall-Sundrum model to be relevant in the context of the hierarchy problem in particle physics, the required ALP/photon coupling turns out to 
be quite large. This implies that light and ultralight ALPs are strongly disfavoured in both the cases we consider - when the gauge field is in the bulk and when it is confined to the 
brane. Restrictive bounds have been estimated on the masses for heavy ALPs, especially in the brane case. We have also shown how the ALP DM scenario is the most strongly constrained in
this context.\\

The tension between the five dimensional flat RS model, ALPs and observational data is quite apparent. The promise of RS model to resolve the gauge hierarchy/fine-tuning problem turns out to be 
questionable in the light of the observed ALP/photon phenomenology - for the bulk case, the ALP mass has to be $m_{a} \gtrsim 0.1$ [GeV] while masses up to a few TeV
are constrained for gauge fields on the brane. After completion of this work we noticed an interesting study on axion-photon couplings in the 
context of $10$ dimensional heterotic string theory to test the viability of different classes of heterotic string models \cite{Agrawal2024}. Our work here can be 
extended to various generalised warped geometry scenarios in the context of multiple warped geometry models as well as branes with non-vanishing curvature.
We plan to look into this and report the outcome in a future work.

%%%%%%%%%%%%%%%%%%%%%%%%%%%%%%%%%%%%%%%%%%%%%%%%%%%%%%%%%%%%%%%%%%%%%%%%%%%%%%%%%%%%%%%%
\section*{Acknowledgements}
SH would like to thank Tanmoy Kumar and Ciaran O'Hare for helpful clarifications and inputs. The datasets and codes have been primarily adapted from 
\cite{CJHAxionLimits} and modified appropriately to derive the relevant bounds. SH would also like to acknowledge the KVPY fellowship provided by the DST, India.

\bibliographystyle{elsarticle-num}
\bibliography{Bibliography}

\end{document}